\documentclass[%
 reprint,
superscriptaddress,
 amsmath,amssymb,
 aip,
]{revtex4}

\usepackage{graphicx}
\usepackage{dcolumn}
\usepackage{bm}
\usepackage{color}
\usepackage{float}

\begin{document}

\hyphenpenalty=500
\title{Exciton Transport Coherent with Upper and Lower Polaritons}
\author{Jingyu Liu$^{1}$, Yao Yao$^{1,2}$}
\address{$^1$ Department of Physics, South China University of Technology, Guangzhou 510640, China \\
$^2$ State Key Laboratory of Luminescent Materials and Devices, South China University of Technology, Guangzhou 510640, China}
\email{yaoyao2016@scut.edu.cn}
\date{\today}

\begin{abstract}
In the strong exciton-cavity coupling regime, exciton polaritons play a pivotal role in governing exciton transport. However, the specific contributions of the two branches of polaritons to exciton current remain unexplored. Here, we decompose the exciton current operator into components coherent with the upper and lower polaritons. Subsequently, we employ the quantum master equation to investigate the exciton current coherent with these two polariton branches. Our findings reveal that in the blue detuning regime, where the optical frequency exceeds the molecular excitation energy, the contribution of the exciton current coherent with the lower polariton is more pronounced. Conversely, in the red detuning regime, where the optical frequency is lower than the molecular excitation energy, the contribution of the exciton current coherent with the upper polariton prevails. We elucidate two critical factors influencing these contributions: firstly, the ratio of excitons to photons in the polariton branches. Due to the smaller decay rate of molecules compared to that of the cavity, the polariton branch with a higher proportion of excitons contributes more to coherent exciton current. Secondly, the energy gap between the polariton branch and the dark states; a smaller energy difference in a branch provides a larger contribution to coherent exciton current.
\end{abstract}
\maketitle


\section{\label{sec:Introduction}Introduction}

Exciton transport plays a pivotal role in various physical phenomena, spanning from natural photosynthesis to optoelectronic applications such as organic semiconductors \cite{PRB2007agranoich,May2011,NM2015orgiu,AM2020hou} and solar energy harvesting \cite{ARPC2009Cheng,NC2011Scholes,NC2014Oreilly,JCP2015Killoran}. Traditionally, the transport of excitons in molecular materials has been limited by intrinsic dynamic disorder and considerable static disorder, resulting in short-range diffusion with diffusion lengths typically less than $10\,\mathrm{nm}$ \cite{book2013fret,AM2020hou}. However, recent advancements in microcavity engineering and photonic crystal fabrication have opened up new avenues for enhancing exciton transport by leveraging strong light-matter interactions to form hybrid states known as polaritons \cite{NM2015orgiu,PRL2015Scha,PRL2015Feist,AC2016non,AC2017energy,CheSci2018Ribeiro,georgiou2018control,PRL2021Chavez,AS2022tuning,NM2023Bala}.  The strong coupling is a relative concept, only achieved when the rate of energy transfer between light and matter exceeds the system's decay rate, leading to the formation of stable polaritons \cite{ebbesen2016hybrid,NP2017room}. Polaritons, characterized by their smaller effective mass and easier controllability, serve as an ideal platform for manipulating light-matter interactions at the nanoscale \cite{RMP2010Deng,Nature2009Amo,NM2016Sanvitto}, as well as a promising candidate for Bose–Einstein condensation \cite{ReMoP2010hui,NatMa2014plumhof}.

The enhancement of exciton transport induced by the strong coupling between exciton and cavity modes is attributed to the delocalized polaritonic modes, which can efficiently bypass the molecular array \cite{PRL2015Feist}. In realistic systems, various factors can affect exciton transport, such as the broadening resulting from disorder \cite{AM2020hou,PRL2021disorder,JCP2022effects}, intra- and inter-molecular vibrations \cite{JCP2019Liu,PRB2018Organic,JCP2023vibrational,JCP2020liu}, and the Q-factor of the microcavity \cite{AS2022tuning}. Additionally, strong light-matter coupling has been employed to investigate the transport properties of charged particles \cite{ACSNano2020conductivity,PRL2020Kra}. In a standard Tavis-Cummings model \cite{PR1969TC}, coupling N dipoles to a single-mode cavity results in a highest-energy upper polariton state, a lowest-energy lower polariton state and N-1 dark states devoid of photon components. Regarding the influence of dark states on exciton transport, reference \cite{AS2022tuning} finds that the more delocalized dark states facilitate the transport of exciton polaritons. Specifically, they achieved this by increasing the cavity's Q-factor to reduce the bandwidth of the dark state band, thereby increasing its delocalization. The enhanced delocalization of dark states leads to the faster transport of polaritons through their conversion with polariton states. While the mechanism of dark states' action has been reasonably elucidated, investigations into the contributions of coherent exciton transport with upper and lower polariton states remain unexplored.

In this article, we consider a structure similar to the source-bridge-drain configuration commonly used for handling electron transport in semiconductor quantum dots \cite{book1995Datta}. The Hamiltonian describing the central molecule and the microcavity region encompasses exciton, cavity, and exciton-cavity coupling terms. We incorporate exciton decay, exciton dephasing, and cavity photon decay to capture the dynamics of the system. By employing quantum master equation, we investigate exciton currents coherent with both upper and lower polaritons, revealing insights into how detuning and molecular decay rate influence each polariton branch's contribution. In the regime of blue detuning , where the optical frequency exceeds the molecular excitation energy, we observe a pronounced contribution from the exciton current coherent with the lower polariton. Conversely, in the red detuning regime, characterized by an optical frequency lower than the molecular excitation energy, the contribution of the exciton current coherent with the upper polariton predominates. We found two critical factors determine the relative contributions of polariton branches to coherent exciton current: the ratio of excitons to photons within each branch and the energy gap between the polariton branches and the dark states.

The rest of this paper is organized as follows: Section \ref{sec:1} elaborates on the system and methodology. Section \ref{sec:2} calculates the exciton current coherent with the upper (lower) polariton.  Conclusions and discussions are drawn in Sec. \ref{sec:3}.

\section{\label{sec:1}System and Methodology}

\begin{figure}
\begin{center}
\includegraphics[width=0.45\textwidth]{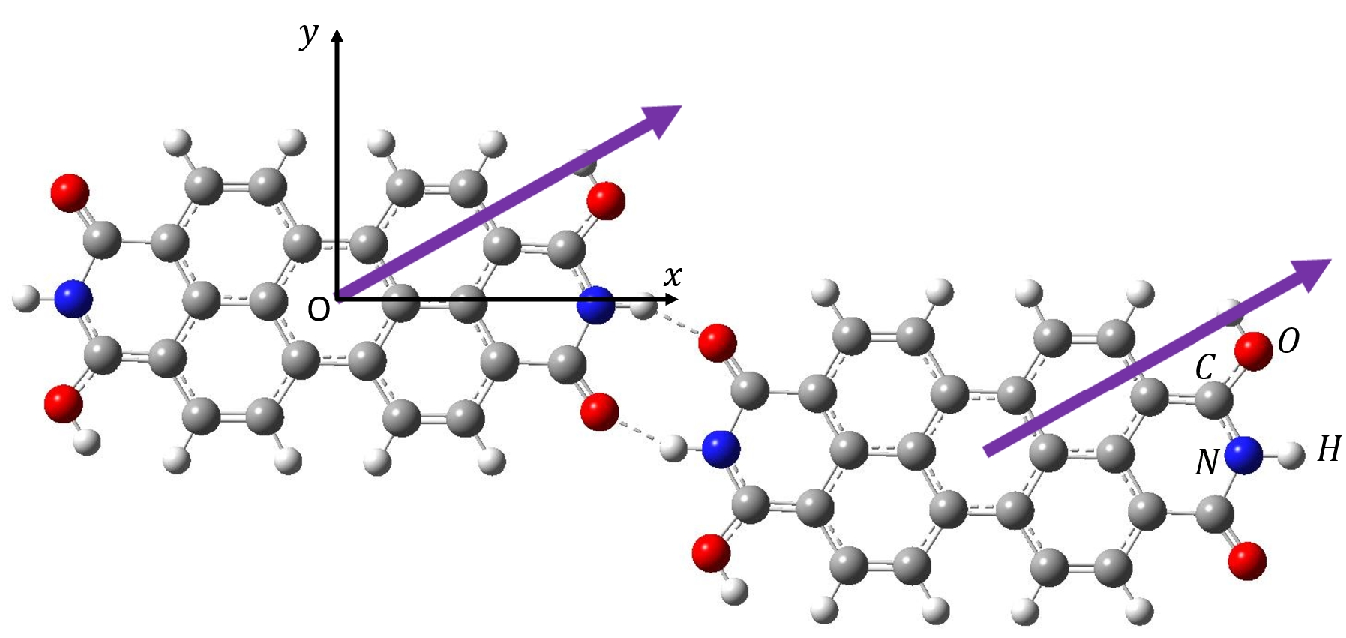}
\caption{Two perylene diimide molecules interconnected through hydrogen bonding. Each molecule is simplified as an electric dipole moment.}
\label{fig:sketch1}
\end{center}
\end{figure}

\begin{figure}
\begin{center}
\includegraphics[width=0.45\textwidth]{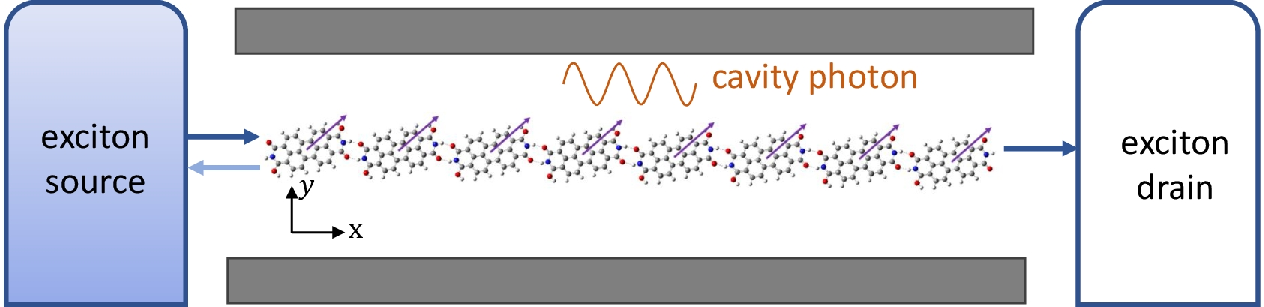}
\caption{Schematic of the source-molecule-bridge-drain structure. The molecular aggregates coupled to the microcavity are connected at both ends to the exciton source and exciton drain. Each molecule is considered as an electric dipole moment.}
\label{fig:sketch2}
\end{center}
\end{figure}

Our research subject is an aggregated form of perylene diimide molecules ($\rm{C_{24}H_{12}O_4N_2}$). The lateral connections between adjacent molecules occur through hydrogen bonding, as illustrated in Fig. \ref{fig:sketch1}. The first excitation energy of a single perylene diimide molecule is 2.18 eV, corresponding to a transition dipole moment of (2.227, 1.085, 0) $\rm{Au}$. The distance between molecules is $1.42 {\rm{nm}}$, from which the dipole-dipole interaction can be calculated as $J=-0.0074 {\rm{eV}}$. Due to its second excitation energy measuring 2.85 eV, significantly higher than the first excitation energy, this molecule can be reasonably considered a two-level system within a specific range of the  optical spectrum. Substituting each molecule with a dipole moment resulted in the schematic representation of the molecular aggregates depicted in Fig. \ref{fig:sketch2}.

To describe exciton transport within the molecular aggregates, we considered a structure similar to the source-bridge-drain configuration commonly used for handling electron transport in semiconductor quantum dots, as illustrated in Fig. \ref{fig:sketch2}. We coupled the perylene diimide polymer to a single-mode microcavity. Excitation energy is pumped into the system from the left exciton source at a rate of $\gamma_{L}$ and traverses through the system, reaching the right exciton drain at a rate of $\gamma_R$. Under the Heitler-London approximation \cite{PR2015schroter} and the rotating wave approximation \cite{PRL2015Feist}, the Hamiltonian for the central molecule and the microcavity region can be described as:
\begin{equation}
\begin{aligned}
& H=H_{\mathrm{e}}+H_{\mathrm{c}}+H_{\mathrm{e}-\mathrm{c}}, \\
& H_{\mathrm{e}}=\sum_{n=1}^N \varepsilon a_{n}^{\dagger} a_{n}+\sum_{n=1}^{N-1} J(a_{n}^{\dagger} a_{n+1}+a_{n+1}^{\dagger} a_{n}), \\
& H_{\mathrm{c}}=\omega_{\mathrm{c}} c^{\dagger} c, \\
& H_{\mathrm{e}-\mathrm{c}}=\sum_{n} g_{n}\left(a_{n}^{\dagger} c+c^{\dagger} a_{n}\right),
\end{aligned}
\end{equation}
where the Hamiltonian $H$ encompasses the following constituents: $H_{\rm{e}}$ represents the exciton component, $H_{\rm{c}}$ depicts a single-mode cavity, and $H_{\rm{e-c}}$ accounts for the exciton-cavity coupling. The operators $a_n^\dagger$ and $a_n$ create and annihilate excitations of the $n$th molecule, possessing an energy $\varepsilon$, respectively. The parameter $J$, representing the dipole-dipole interaction \cite{PR2015schroter}, denotes the nearest-neighbor excitation transfer interaction. $c^{\dagger}$ and $c$ act as photon creation and annihilation operators with energy $\omega_c$, respectively. It's worth mentioning that the current model differs from the Tavis-Cummings (TC) \cite{PR1969TC} or Dicke model \cite{PR1954Dicke} primarily because of the additional excitation transfer interaction. We assume that the polarization direction of the electric field in this single-mode cavity is along the x-axis direction. Consequently, the coupling between the molecular dipole moment and the electric field becomes a unified numerical value $g_n=g$.

Furthermore, we incorporate exciton decay ($\gamma_d$) attributable to spontaneous emission, exciton dephasing ($\gamma_p$) caused by interaction with the continuous phonon bath, and the cavity photon decay ($\kappa$) due to leakage through the mirrors. The dynamics of the entire system is thus governed by the master equation
\begin{equation}
\begin{aligned}
\frac{\partial \rho}{\partial t}=& -\frac{i}{\hbar} [H,\rho]\!+\!\gamma_{L}\mathcal{L}_{a_{1}^{\dagger}}(\rho)\!+\!\gamma_{L}(\bar{n}_{L}\!+\!1) \mathcal{L}_{a_{1}}(\rho)\!+\!\gamma_{L}\mathcal{L}_{a_{N}}(\rho)\\
&+\sum_{n=1}^{N}\left[\gamma_{d}\mathcal{L}_{a_{n}^{\dagger}}(\rho)\!+\!\gamma_{p}\mathcal{L}_{a_{n}^{\dagger}a_n}(\rho)\right]\!+\!\kappa\mathcal{L}_{c}(\rho),
\end{aligned}
\end{equation}
where the operator $\mathcal{L}_a(\rho)=a\rho a^\dagger-\frac{1}{2}\{\rho,a^\dagger a\}$ represents the standard Lindblad superoperator.  The first term on the right-hand side represents the unitary evolution of the system, while the second and third terms denote the injection of excitons from the exciton source and the reflection of excitons back into the exciton source, respectively. Here, $\bar{n}_{L}$ represents the average number of excitons in the exciton source. The fourth term signifies the current of excitons out to the exciton drain. The fifth and sixth terms correspond to the decay of excitons and the leakage of cavity photons, respectively. By utilizing the master equation, we can compute the density matrix at steady-state and calculate the exciton current.

Our subsequent calculations will be performed in the subspace of the excitation number $\sum_{n=1}^N a_{n}^{\dagger} a_{n}+c^\dagger c=\{0,1\}$. Within this subspace, our defined basis includes $|0\rangle$, $|n\rangle$ and $|c\rangle$, where $|0\rangle$ represents the vacuum state, $|n\rangle$ represents the excited state of the $n$th molecule, and $|c\rangle$ denotes the single-photon state. To maintain the validity of this low-excitation number, we set the excitation rate of the system $\gamma_L$ to be much smaller than the leakage rate of cavity photons $\kappa$, ensuring that the system remains at a low excitation number.

To explore the contributions of upper and lower polaritons to the exciton current, we define the exciton current operator flowing into the $N$th molecule. The excitation number operator for the $N$th molecule is $\mathrm{X}_{N}=a_N^\dagger a_N$. The time derivative of its average value is calculated using the commutator with the Hamiltonian,
\begin{equation}
\frac{{\rm{d}}}{{\rm{d}}t}\langle\mathrm{X}_{N}\rangle_{t}=\frac{-i}{\hbar}\left\langle\left[\mathrm{X}_{N}, H\right]\right\rangle_{t},
\end{equation}

where $\langle \mathrm{X}_{N}\rangle_{t}=\mathrm{Trace}\{\rho(t)\mathrm{X}_{N}\}$  represents the average value of the excitation number operator for the $N$th molecule at time $t$. Thus, the exciton current operator flowing into the $N$th molecule is obtained as
\begin{equation}
\begin{aligned}
\hat{j}_\mathrm{full}=&\frac{-i}{\hbar}\left[\mathrm{X}_{N}, H\right]\\
                     =&\frac{-i}{\hbar}J(a_N^\dagger a_{N-1}\!+\!a_{N-1}^\dagger a_{N})\!+\!\frac{-i}{\hbar}g(a_N^\dagger c\!+\!c^\dagger a_N),
\end{aligned}
\end{equation}

where the first term in the equation represents the exciton current transmitted through excitation transfer interactions, the second term signifies the exciton current transmitted through the exciton-cavity coupling, denoted as $\hat{j}_{c}$. Previous studies \cite{PRL2015Scha,PRL2015Feist} have pointed out that in weakly coupled exciton-cavity systems, excitation transfer interactions primarily contribute to exciton transport, while in strongly coupled systems, polaritons dominate. We separately calculated cases with $J\neq 0$ and $J=0$, as shown in Fig. \ref{fig3}. It is evident that the total exciton current $\langle\hat{j}_\mathrm{full}\rangle$ calculated with strong exciton-cavity coupling ($J\neq 0$) shows minimal deviation from $\hat{j}_{c}$ calculated with $J=0$. Therefore, for subsequent calculations, to conveniently explore the contributions of upper and lower polaritons to the exciton current, we neglect excitation transfer interactions, i.e., $J=0$.

Under strong coupling, excitons and photons can form polaritons. Conversely, we can also express the creation and annihilation operators of photons in terms of upper and lower polaritons in the exciton current operator:
\begin{equation}\label{exciton current}
\begin{aligned}
\hat{j}_{c} & =\frac{-i}{\hbar} (a_N^\dagger c+c^\dagger a_N) \\
            & =\frac{-i}{\hbar} g_n(C_{\mathrm{U}}a_N^\dagger a_{\mathrm{UP}}\!-\!C_{\mathrm{U}}^{*} a_{\mathrm{UP}}^\dagger a_N)\\
            &  +\frac{-i}{\hbar} g_n(C_{\mathrm{L}}a_N^\dagger a_{\mathrm{LP}}\!-\!C_{\mathrm{L}}^{*} a_{\mathrm{LP}}^\dagger a_N)\\
            & =\hat{j}_{\mathrm{up}}+\hat{j}_{\mathrm{low}},
\end{aligned}
\end{equation}
where  $c=C_{\mathrm{U}}a_{\mathrm{UP}}+C_{\mathrm{L}}a_{\mathrm{LP}}$, $c^\dagger=C_{\mathrm{U}}^{*}a_{\mathrm{UP}}^\dagger+C_{\mathrm{L}}^{*}a_{\mathrm{LP}}^\dagger$. The energy of upper and lower polaritons can be referred to in Fig. \ref{fig4}. According to Equation \ref{exciton current}, it is evident that $\hat{j}_{\mathrm{up}}$ ($\hat{j}_{\mathrm{low}}$) represents coherent transport between the upper (lower) polariton state and other states, particularly the dark states, as the terms resembling $C_{\mathrm{U}}a_N^\dagger a_{\mathrm{UP}}$ correspond to non-diagonal elements calculated in the polariton representation \cite{JPCL2013does}. With the exciton current operators coherent with upper and lower polaritons, we can now separately analyze their contributions.

\section{\label{sec:2}Numerical Results}

In the numerical simulation, we choose system parameters approximately corresponding to J aggregates at room temperature \cite{PRL2015Feist}: $\gamma_d^{-1}=600\,\mathrm{fs}$, $\gamma_p^{-1}=25\,\mathrm{fs}$, and $\bar{n}_{L}=2$. Microcavities with metallic mirrors ($Q\sim 10$) exhibit photon decay times of only a few tens of femtoseconds \cite{ACSPho2018Herrera}. We set $\kappa^{-1}=50\,\mathrm{fs}$, a duration shorter than the molecular decay time. To ensure the validity of the single-excitation approximation, we set the pumping rate to $\gamma_{L}^{-1}=\gamma_{R}^{-1}=100\,\mathrm{fs}$.
\begin{figure}
\begin{center}
\includegraphics[width=0.45\textwidth]{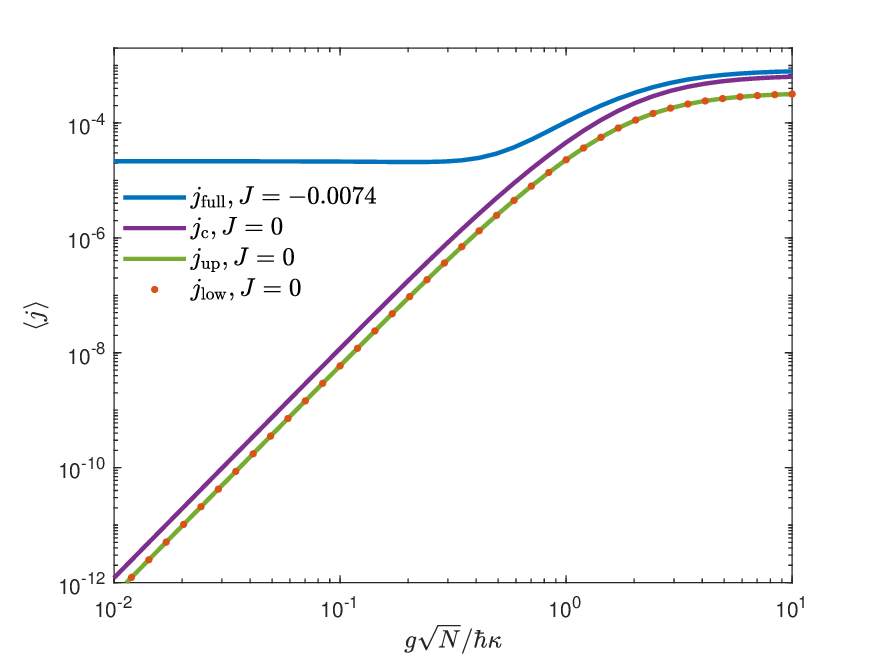}
\caption{Steady-state exciton current as a function of the collective exciton-cavity coupling $g\sqrt{N}$ in the case of zero detuning $\delta=\omega_c-\varepsilon=0$. The blue line represents the total exciton current for $J=-0.0074 {\rm{eV}}$. The purple, green, and red lines correspondingly depict the total exciton current, exciton current coherent with upper-polariton, and exciton current coherent with lower-polariton for $J=0$.  The number of molecules $N=10$, other parameters $\omega_c=\varepsilon=2.18\,\mathrm{eV}$, $\gamma_d^{-1}=600\,\mathrm{fs}$, $\gamma_p^{-1}=25\,\mathrm{fs}$, $\bar{n}_{L}=2$, $\gamma_{L}^{-1}=\gamma_{R}^{-1}=100\,\mathrm{fs}$.}
\label{fig3}
\end{center}
\end{figure}

We initially investigate the scenario of zero detuning, $\delta=\omega_c-\varepsilon=0$, as depicted in Fig. \ref{fig3}. It can be observed that the steady-state exciton currents coherent with upper and lower polaritons are indistinguishable. This uniformity arises because the upper polariton state $|\mathrm{UP}\rangle=\frac{\sqrt{2}}{2}\sum_{n=1}^{N}\frac{1}{\sqrt{N}}|n\rangle +\frac{\sqrt{2}}{2}|c\rangle$ and the lower polariton state $|\mathrm{LP}\rangle=\frac{\sqrt{2}}{2}\sum_{n=1}^{N}\frac{1}{\sqrt{N}}|n\rangle -\frac{\sqrt{2}}{2}|c\rangle$  have the same proportions of photons and excitons.
\begin{figure}
\centering
\includegraphics[width=0.45\textwidth]{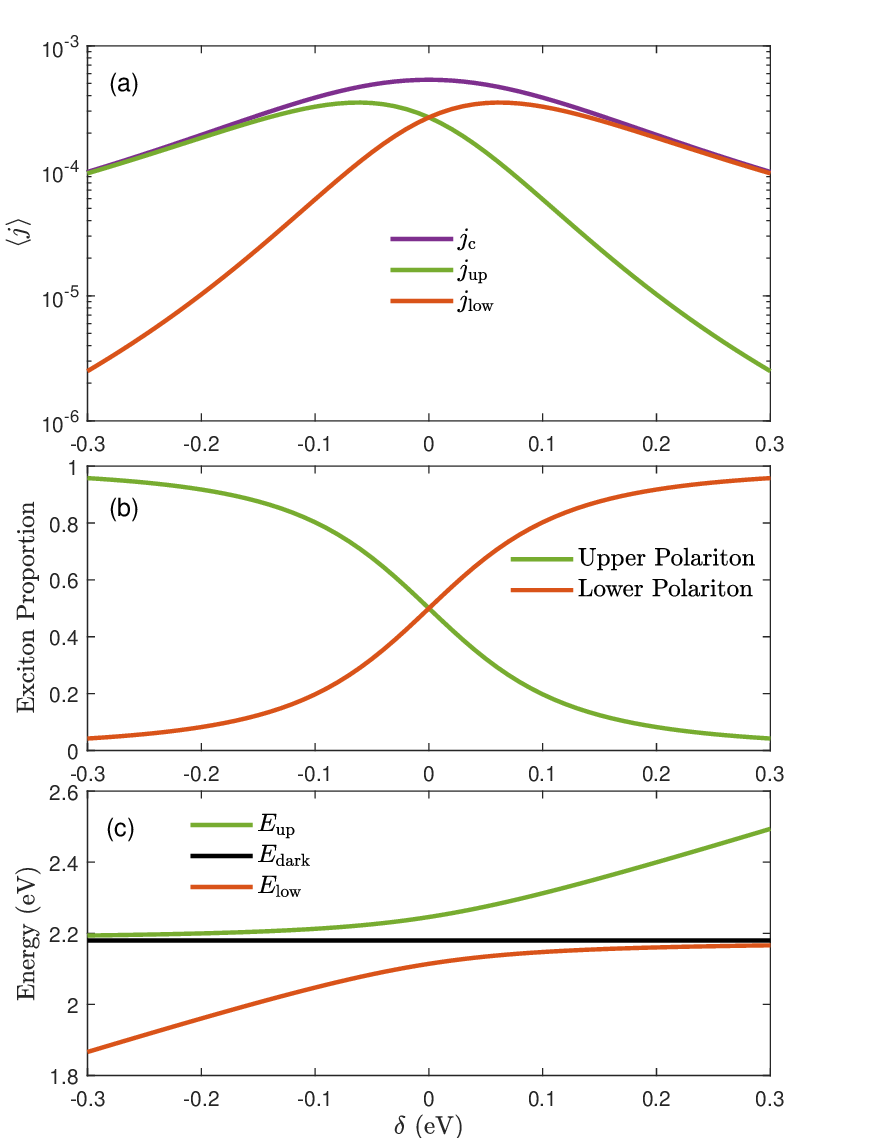}
\caption{In the strong exciton-cavity coupling regime: (a) Steady-state exciton current variation with detuning $\delta=\omega_c-\varepsilon$. (b) The exciton proportion from the upper-polariton state (green) and lower-polariton state (red) as a function of detuning. (c) Energy variation of the upper polariton state (green), dark states (black), and lower polariton state (red)  with respect to detuning. The number of molecules $N=10$, other parameters $\varepsilon=2.18\,\mathrm{eV}$, $J=0\,\mathrm{eV}$, $\gamma_d^{-1}=600\,\mathrm{fs}$, $\gamma_p^{-1}=25\,\mathrm{fs}$, $\bar{n}_{L}=2$, $\gamma_{L}^{-1}=\gamma_{R}^{-1}=100\,\mathrm{fs}$.}
\label{fig4}
\end{figure}

Through varying the optical frequency, we observed that the total exciton current corresponding to zero detuning is a maximum, as shown in Fig. \ref{fig4}(a), which is consistent with the conclusions in Ref. \cite{PRL2015Feist}.  When the detuning is negative (positive), i.e., in the red detuning (blue detuning) regime,  it can also be observed that upper (lower) polaritons dominate the transport. Furthermore, the contributions of coherent exciton currents from upper (lower) polaritons gradually decrease (increase) with increasing detuning, as shown in Fig. \ref{fig4}(b). To explain this phenomenon, we identified two potentially crucial factors. The first factor is related to the proportion of excitons and photons in polaritons. In Fig. \ref{fig4}(b), we observed that the exciton proportion in upper (lower) polariton gradually increases (decreases) with increasing detuning. The decay rate of the molecule is much smaller than that of the photon, $\gamma_d \ll \kappa$. Therefore, we speculate that the rapid decay rate of the photon component will result in a polariton branch with a higher proportion of photons having a smaller coherent exciton current in the two branches. The second factor is related to energy gap, as shown in Fig. 4(c). With increasing detuning, the gap between upper (lower) polaritons and dark states gradually increases (decreases). Ref. \cite{JCP2020liu} found that the contributions of upper and lower polaritons are strongly correlated with the energy gap between them and the dark states, indicating that the closer they are to the dark states, the greater the contribution. Our results align with those of Ref. \cite{JCP2020liu}. This conclusion is also reasonable, as a smaller energy difference usually implies a higher probability of transition.
\begin{figure}
\centering
\includegraphics[width=0.45\textwidth]{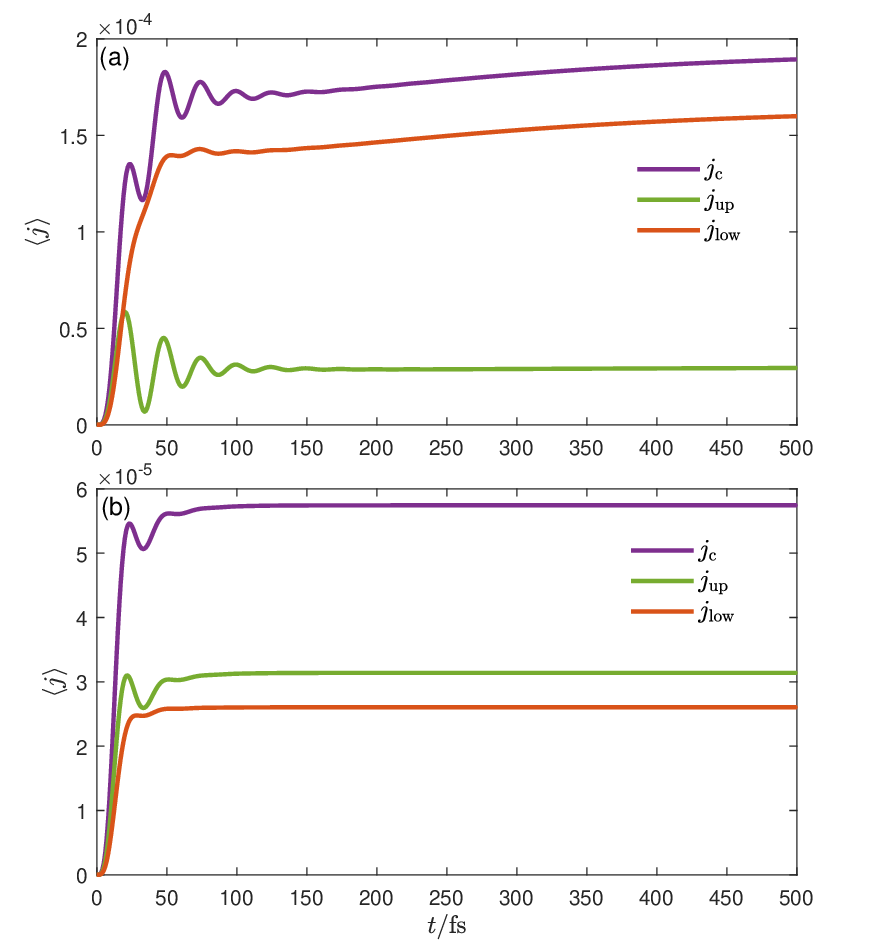}
\caption{The temporal evolution of exciton current under strong coupling $g\sqrt{N}/\hbar \kappa=5$. The initial state of the system is chosen to be the vacuum state. (a) The cavity decay time $\kappa^{-1}=50\,\mathrm{fs}$ is faster than the molecular decay time $\gamma_d^{-1}=600\,\mathrm{fs}$. (b) The molecular decay time $\gamma_d^{-1}=8\,\mathrm{fs}$ is faster than the cavity decay time $\kappa^{-1}=50\,\mathrm{fs}$. The number of molecules $N=10$, other parameters $\varepsilon=2.18\,\mathrm{eV}$, $\omega_c=2.28\,\mathrm{eV}$, $\delta=0.1\,\mathrm{eV}$, $J=0\,\mathrm{eV}$, $\gamma_d^{-1}=600\,\mathrm{fs}$, $\gamma_p^{-1}=25\,\mathrm{fs}$, $\bar{n}_{L}=2$, $\gamma_{L}^{-1}=\gamma_{R}^{-1}=100\,\mathrm{fs}$.}
\label{fig5}
\end{figure}

To verify the first factor, we calculated the temporal evolution of the exciton current under different molecular decay rates. Fig. \ref{fig5}(a) depicts the exciton current for a molecular decay rate of $\gamma_d=\frac{1}{600\,\mathrm{fs}}$. We found that at steady state, the exciton current coherent with upper polaritons is smaller than that coherent with lower polaritons. When we increased the molecular decay rate to be sufficiently fast, $\gamma_d=\frac{1}{8\,\mathrm{fs}}$, we observed that the exciton current coherent with upper polaritons becomes larger than that coherent with lower polaritons. In other words, by increasing the molecular decay rate, we reduced the coherent exciton current of the lower polariton, which has a higher proportion of excitonic components. However, the actual molecular decay rate is much slower than the decay rate of photons \cite{PRL2015Feist}. Therefore, the coherent exciton current associated with the upper (lower) polariton is related to the ratio of excitonic and photonic components, and the polariton branch with a higher excitonic ratio exhibits a larger coherent exciton current.

\section{\label{sec:3}Conclusions and Discussion}

Our investigation focuses on the coherent exciton current associated with upper and lower polaritons in the strong coupling regime. We found that under zero detuning, the steady-state exciton currents coherent with upper and lower polaritons exhibit remarkable similarity. This uniformity stems from the balanced proportions of photons and excitons in both polariton states. By varying the detuning, the total exciton current peaks at zero detuning. Furthermore, in the blue detuning scenario, where the optical frequency exceeds the molecular excitation energy, the contribution of the exciton current coherent with the lower polariton becomes more pronounced. Conversely, in the red detuning regime, where the optical frequency is lower than the molecular excitation energy, the contribution of the exciton current coherent with the upper polariton dominates. We identified two key factors driving these observations. Firstly, the ratio of excitons to photons in the polariton branches plays a pivotal role, with the branch possessing a higher exciton proportion contributing more to coherent exciton current. This is attributed to the slower decay rate of molecules compared to that of the cavity. Secondly, the energy gap between the polariton branch and the dark states significantly influences the contribution to coherent exciton current, with a smaller energy difference leading to a larger contribution. Temporal evolution analyses under varying molecular decay rates further underscored the importance of the ratio of excitons to photons in the polariton branches for determining the contributions to exciton current.

In conclusion, our study elucidates the roles of upper and lower polaritons in governing exciton transport processes. Although our results indicate that the exciton currents coherent with upper polaritons and lower polaritons are equal when the blue detuning equals the red detuning, in an actual system, we speculate that under equal detuning conditions, the red detuning is more conducive to exciton transport than the blue detuning. This is because excitons scatter more readily into lower polariton states with lower energy through vibrational relaxation or exciton-exciton interactions \cite{APL2010longitudinal,PRB2002polariton,NRP2022polariton}. Consequently, the proportion of lower polariton states tends to surpass that of upper polariton states, thereby suggesting a more effective utilization of red detuning in harnessing polaritons. The identified factors provide insights into designing and optimizing systems operating in the strong coupling regime, facilitating enhanced coherent exciton current in such setups. By deeply understanding the role of upper and lower polaritons in exciton transport, we can not only enhance our comprehension of highly efficient exciton transport mechanisms but also optimize the utilization of these physical processes for designing optoelectronic devices with superior performance.

\noindent{\bf DATA AVAILABILITY}
\par The data that support the findings of this study are available from the corresponding author upon reasonable request.

\begin{acknowledgments}
The authors gratefully acknowledge support from Science and Technology Planning Project of Guangzhou (Grant No.~202201010696).
\end{acknowledgments}

\section*{REFERENCES}
\bibliography{transport.bbl}

\end{document}